\documentclass[runningheads]{llncs}
\usepackage{graphicx}
\usepackage{balance}  

\usepackage{url,afterpage,amssymb,amsfonts}
\usepackage{enumitem,makecell}
\usepackage{booktabs}
\usepackage{multirow}
\usepackage{subfig}
\usepackage{graphicx}
\usepackage{amsmath}
\usepackage{verbatim}
\usepackage{algpseudocode}
\usepackage{color}
\usepackage{colortbl}
\graphicspath{{./figs/}}
\usepackage[linesnumbered,noend]{algorithm2e}

\usepackage{xcolor}
\usepackage{listings}
\usepackage{soul}
\usepackage{subfig}
\usepackage{url,afterpage,amssymb,amsfonts}
\usepackage{enumitem,makecell}
\usepackage{color}
\usepackage{colortbl}
\usepackage{float}
\usepackage{wrapfig}

\definecolor{mGreen}{rgb}{0,0.6,0}
\definecolor{mGray}{rgb}{0.5,0.5,0.5}
\definecolor{mPurple}{rgb}{0.58,0,0.82}
\definecolor{backgroundColour}{rgb}{0.95,0.95,0.92}

\newcolumntype{s}{>{\columncolor[gray]{0.8}}r}
\newcolumntype{t}{>{\columncolor[gray]{0.8}}c}

	{\end{list}}

\lstdefinestyle{CStyle}{
	backgroundcolor=\color{backgroundColour},
	commentstyle=\color{mGreen},
	keywordstyle=\color{magenta},
	numberstyle=\tiny\color{mGray},
	stringstyle=\color{mPurple},
	basicstyle=\scriptsize,
	breakatwhitespace=false,
	breaklines=true,
	captionpos=b,
	keepspaces=true,
	numbers=left,
	numbersep=5pt,
	showspaces=false,
	showstringspaces=false,
	showtabs=false,
	tabsize=2,
	language=C
}

\graphicspath{{figs/}}
\def\BibTeX{{\rm B\kern-.05em{\sc i\kern-.025em b}\kern-.08em
		T\kern-.1667em\lower.7ex\hbox{E}\kern-.125emX}}

\algrenewcommand\algorithmicindent{0.65em}

\def\sssp{\mbox{\sc sssp}}

\begin{document}
\title{An Adaptive Load Balancer For Graph Analytics on GPUs}
%
%
\author{Vishwesh Jatala\inst{1,2} \and
Loc Hoang\inst{1,3} \and
Roshan Dathathri\inst{1,3} \and
Gurbinder Gill\inst{1,3} \and
V.~Krishna Nandivada\inst{4,5} \and
Keshav Pingali\inst{1,3}
}
\authorrunning{V. Jatala et al.}
%
\institute{The University of Texas at Austin, Texas, United States \and
\email{vishwesh.jatala@austin.utexas.edu} \and 	
\email{\{loc,roshan,gill,pingali\}@cs.utexas.edu} \and
Indian Institute of Technology Madras, Chennai, India \and
\email{nvk@iitm.ac.in}}
\maketitle              

\begin{abstract} \label{sec:abstract}
Load-balancing among the threads of a GPU for
graph analytics workloads is difficult because of the
irregular nature of graph applications and
the high variability in vertex degrees,
particularly in power-law graphs. We describe a novel
load balancing scheme to address this problem.
Our scheme is implemented in the IrGL compiler
to allow users to generate efficient load balanced code
for a GPU from high-level sequential programs.
We evaluated several graph analytics applications on up to
16 distributed GPUs using IrGL to compile the code 
and the Gluon substrate for inter-GPU communication.
Our experiments
show that this scheme can achieve an average speed-up of
$2.2\times$  on inputs that suffer from severe load imbalance problems 
when previous state-of-the-art load-balancing schemes
are used.


\end{abstract}

\section{Introduction}
\label{sec:intro}

Graphics processing units (GPUs) have become popular platforms for processing
graph analytical applications~\cite{irgl,gluon,lux,groute,gunrock}. In spite of the
computational advantages provided by GPUs, achieving good performance for
graph analytical applications remains a challenge. Specifically, load balancing
across GPU threads as well as among multiple GPUs is a difficult
problem since many distributed applications execute in bulk-synchronous rounds, and
imbalance among threads within a GPU in a round may cause all threads to wait for
stragglers to complete.

Load balancing in multi-GPU systems is a difficult problem for many reasons. 
The first reason is that the set of vertices to be processed in a computational round is
statically unpredictable and may vary dramatically from round to round.
Therefore, static load balancing techniques do not work well. Another
complication is that most large graphs today are power-law graphs in which
vertex degrees follow a power-law distribution (i.e., a few vertices have
orders of magnitude more neighbors than the rest).  Therefore, simple load
balancing schemes that assign vertices to threads may not perform well.
Finally, good load balancing schemes must account for the architecture of
modern GPUs and hierarchy of threads: thread-blocks, warps, and threads.

Several graph processing frameworks have proposed load balancing
strategies for graph analytics on GPUs~\cite{merrill12,enterprisegpu,gunrock,nasre13datatopology}. 
Most of these strategies involve dynamically partitioning vertices or edges
evenly across the thread blocks, warps, or threads of the GPU.  However, 
they have one or more of these limitations: (i) they do not load balance across 
thread blocks, (ii) they have high memory or computation overheads, or 
(iii) they require high programming effort.

We present an adaptive load balancing strategy called ALB that addresses
load imbalance at runtime. In each computation round, it classifies vertices
based on their degrees since vertex degrees provide an estimate of load imbalance.
Edges of very high degree vertices are evenly assigned across all 
threads in all thread blocks, using a novel cyclic edge distribution strategy 
that accounts for the memory access patterns of the GPU threads. 
All other vertices are evenly distributed across thread blocks, warps,
or threads similar to a prior load balancing scheme~\cite{merrill12}.
We implemented our strategy in the IrGL compiler~\cite{irgl}, which permits
users to write sequential graph analytics programs without knowledge
of GPU architectures.
The generated compiler code inter-operates with the Gluon communication
substrate~\cite{gluon}, enabling it to run on multiple GPUs in a
distributed-memory cluster.

We evaluated the benefits of our approach on a single machine with up to
8 GPUs and on a distributed GPU cluster with up to 16 GPUs.
We compare our approach with other frameworks that support different
load balancing strategies. Our experiments show that our
load balanced code achieves an average speedup of (1) $1.6\times$
on a single GPU and (2) $2.2\times$ on multiple GPUs for many graph applications.
Our load-balanced code achieves an average speedup of $2.2\times$
compared to other third-party frameworks on power-law graphs while incurring
negligible overhead in inputs that do not suffer from heavy load imbalance.





\section{Background on Graph Analytics and GPUs}
\label{sec:background}


\noindent{\textbf{Graph Analytics:}}
A graph consists of vertices, edges, and their associated labels.  Vertex
labels are initialized at the start and updated repeatedly until some global
quiescence condition is reached. Updates to labels are performed by applying an
{\em operator} to {\em active vertices} in the 
graph~\cite{pingali11}.
Push operators read the label of the active vertex and updates the labels of its
immediate neighbors. Pull operators read the labels of the immediate neighbors
of the active vertex and updates the label of the active vertex.
To process a graph on a distributed cluster, the input graph is partitioned
among machines.  Execution occurs in bulk-synchronous parallel (BSP) rounds: in
each round, hosts compute on local partitions and then participate in global
synchronization in which labels of vertices are made consistent.

\noindent{\textbf{GPU Execution:}}
GPUs offer higher memory bandwidth and more concurrency than most CPUs; both
can be exploited for high-performance graph analytics.
A GPU executes multithreaded programs called {\it kernels}. 
A kernel is launched on the GPU with a fixed number of threads.  The CUDA
programming model (used for NVIDIA GPUs) for kernels is hierarchical -- each
kernel executes as a collection of {\it thread blocks\/} 
or cooperative thread arrays (CTA). 
The threads in the CTA are divided into sets called \emph{warps}. The threads
in a warp execute program instructions in a SPMD manner. Once a CTA finishes
its execution, another CTA can be launched.  A kernel ends once all CTAs (thread
blocks) finish.



\begin{figure*}[t]
 \subfloat[\scriptsize{SSSP on rmat25 for different rounds\label{fig:Loadrounds}}]{%
  \includegraphics[scale=0.245]{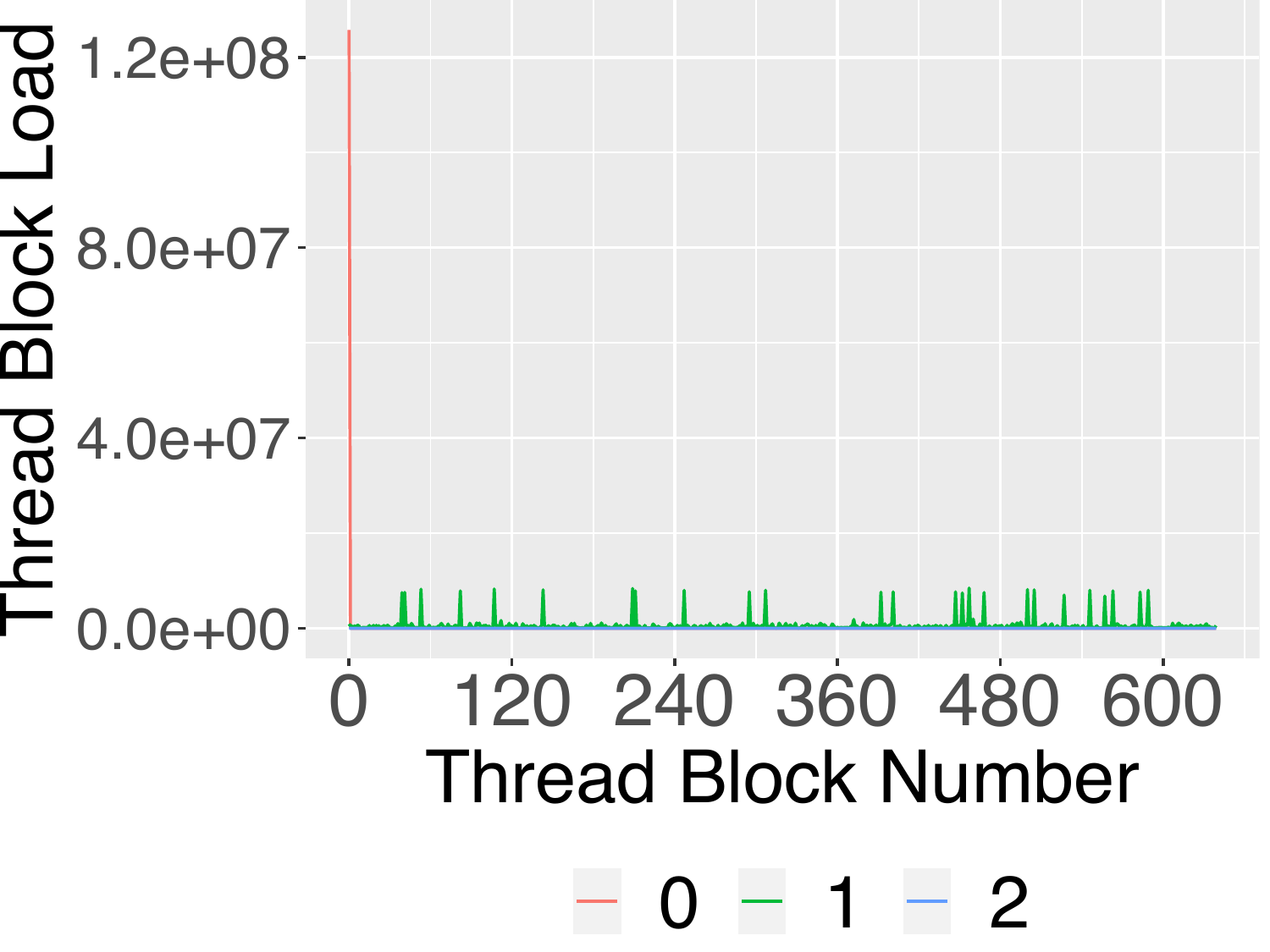}
}
\hspace{7pt}
\subfloat[\scriptsize{BFS on rmat25 and road-USA\label{fig:LoadInput}}]{%
  \includegraphics[scale=0.245]{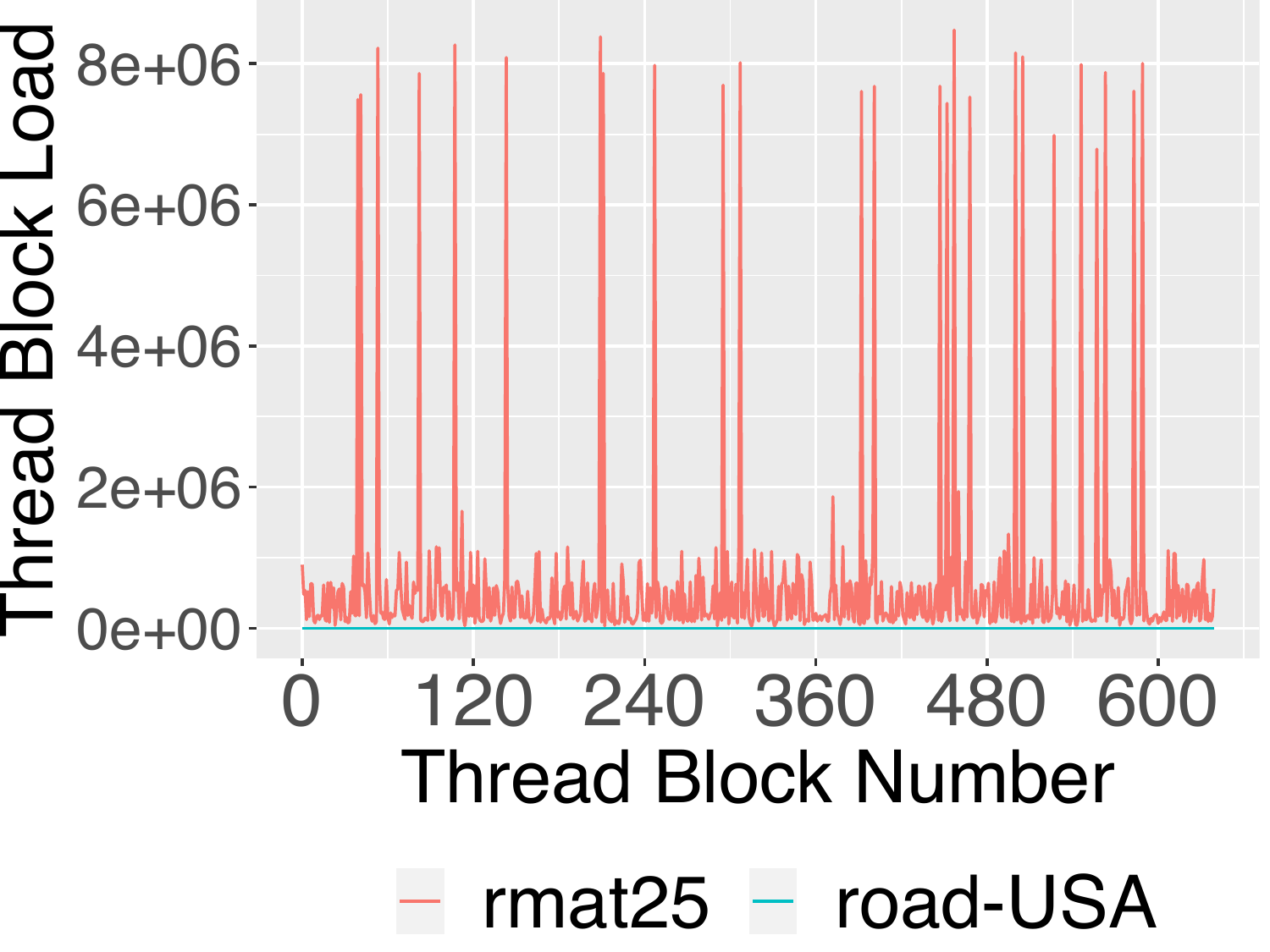}
}%
\hspace{7pt}
\subfloat[\scriptsize{SSSP and PR on rmat25 in $1^{st}$ round\label{fig:LoadApplications}}]{%
\includegraphics[scale=0.245]{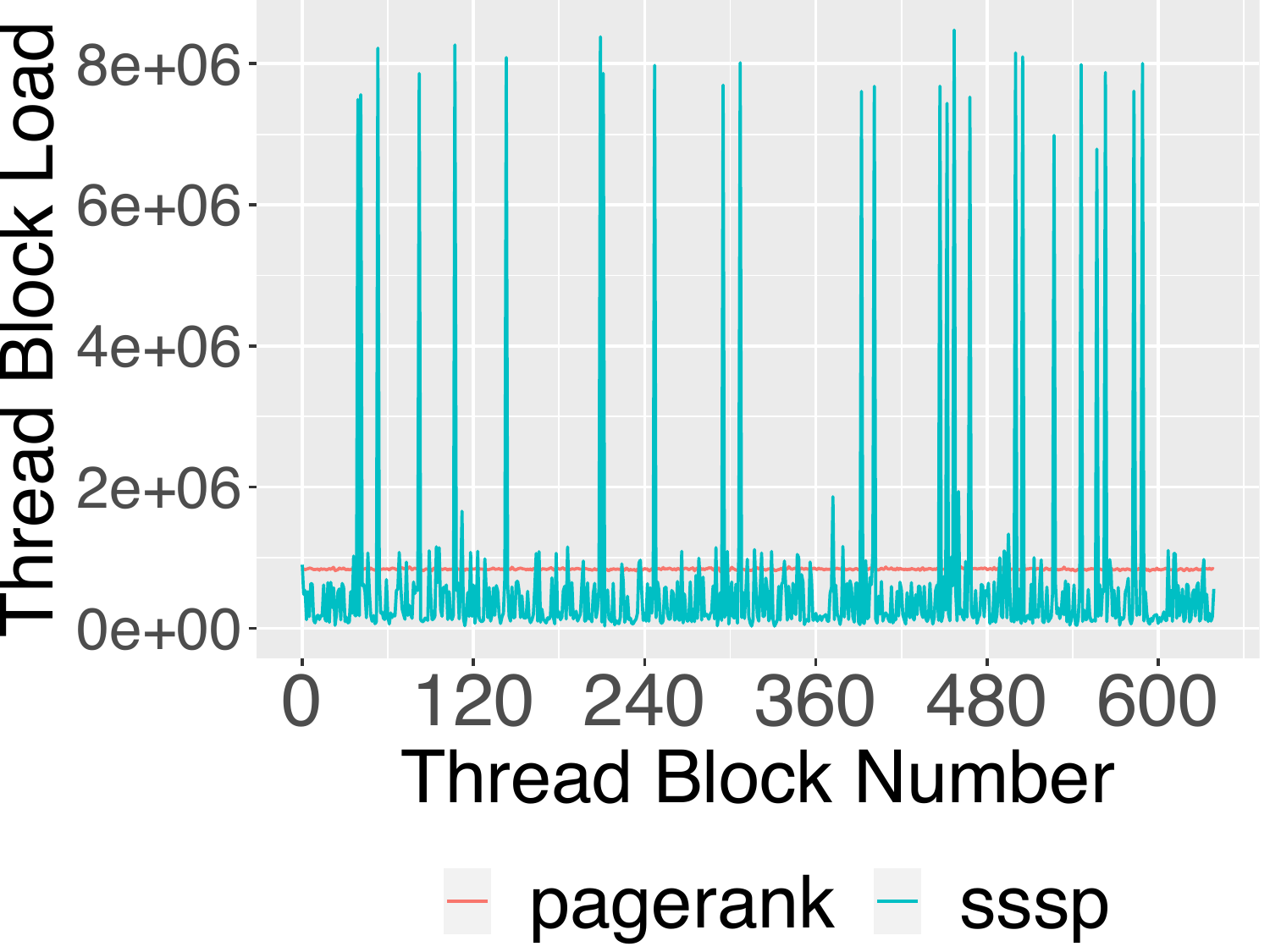}
}
\caption{Thread block load imbalance for various configurations}
\end{figure*}


\section{Existing Load Balancing Schemes}
\label{sec:challenges}


\noindent{\textbf{Vertex/Edge-Based:}} \emph{Vertex-based} load
balancing~\cite{nasre13datatopology} assigns roughly equal numbers of vertices
to GPU threads.  Each thread processes the edges of
vertices assigned to it. This scheme works well if the number of edges
connected to each vertex is roughly the same.  Otherwise, if degree
distribution is skewed like in power-law graphs, it results in severe load
imbalance at an inter-thread and intra-thread block level.
\emph{Edge-based} load balancing~\cite{bcgpus} assigns roughly equal numbers of
edges to each GPU thread, and each thread updates its assigned edges'
endpoints.  This balances the workload, but it needs a graph representation that allows
a thread to quickly access the endpoints and data of an edge from the edge ID
such as coordinate (COO) format or an edge list format, which requires
significantly more space than compressed sparse row (CSR) or compressed sparse
column (CSC) formats.

\noindent{\textbf{Thread-Warp-CTA (TWC):}}
\label{sec:twc}
TWC load balancing~\cite{merrill12} assigns a vertex and its edges to a
thread, a warp, or a thread block (CTA) based on the degree of the
vertex. Each vertex is assigned to one of three
bins (\emph{small}, \emph{medium}, and \emph{large}) based on its degree.
Vertices in the large bin are processed at the thread block level,
vertices in the medium bin at the warp level,
and vertices in the small bin at the thread level.
This scheme load balances all threads within the thread block and does not
need edge endpoints like edge-based balancing.

This scheme may result in load imbalance if degree distributions within a bin vary
significantly, particularly for the large bin which has no upper bound on
degree. To demonstrate this, we measured the load distribution (number of processed
edges) on different thread blocks in the first three rounds of a single-source shortest path
(sssp) computation on the rmat25 graph, using the TWC load-balancing scheme in D-IrGL~\cite{gluon}.
Figure~\ref{fig:Loadrounds} shows that the first two rounds take most of
the execution time, and computational load across thread blocks
is imbalanced.
Thread block imbalance can also vary across different applications for the same
input. We use Figure~\ref{fig:LoadApplications} to illustrate this: sssp
suffers from thread block load imbalance but pagerank does not.
Finally, computational load can be distributed differently across thread blocks
for different inputs.  Figure~\ref{fig:LoadInput} shows the thread block
distribution for bfs on road-USA and rmat25 inputs. Here, bfs exhibits thread
block load imbalance on rmat25 but not on road-USA.

\noindent{\textbf{Vertex Across CTA:}}
Some schemes~\cite{enterprisegpu,gunrock} distribute the edges of a vertex
across thread blocks (CTAs) while using a CSR (or CSC) format.
In the CSR format, only the destination of an edge can be quickly accessed.
To access the source, Enterprise~\cite{enterprisegpu} fixes the source in a round
(they only consider bfs), whereas a load balancing scheme called LB in Gunrock~\cite{gunrock}
finds the source on the fly. Enterprise adds another bin ({\it huge}) to TWC, and
for each vertex in the huge bin,
it processes vertices in rounds (barriers) and
distributes edges of the (source) vertex in a round across threads in all CTAs.
LB assigns edges of all vertices equally among all threads (in all CTAs).
The thread uses binary search to find its source for each assigned edge.
LB uses this distribution for all vertices in all rounds,
so the overheads of binary search might not offset the benefits.
Thus, Enterprise and LB have computation overheads due to
barriers and binary searches.

\section{Adaptive Load Balancer (ALB)}
\label{sec:lb}

\begin{algorithm}[t]
\caption{Inspection Phase of Adaptive Load Balancer}\label{algo:inspect}
\footnotesize
\SetKwFunction{iwtlc}{\textbf{InspectWithTWC}}
\SetKwProg{Fn}{\textbf{procedure}}{:}{}
\Fn{\iwtlc{\textbf{Graph} g, \textbf{Worklist} wl, \textbf{Work} work}}
{
  \ForEach{src in wl}{ \label{line:traverse} 
    degree = getOutDegree(src); \label{line:start} \\
    \If{degree $\geq$ THRESHOLD} { \label{line:degreeCheck}
      work.push(src)  \label{line:push}
    } \Else{
      distribute src edges to thread/warp/CTA based on degree  \label{line:twc}
    } \label{line:end}
  }
}
\end{algorithm}

\vspace{-5pt}
\begin{algorithm}[t]
\caption{Execution Phase of Adaptive Load Balancer}\label{algo:execute}
\footnotesize


\SetKwFunction{iwtlc}{\textbf{LB}}
\SetKwProg{Fn}{\textbf{procedure}}{:}{}
\Fn{\iwtlc{\textbf{Graph} g, \textbf{Worklist} wl, \textbf{Work} work,
\textbf{Worklist} prefixWork}}
{
  edgelist = getMyEdges(work, prefixWork) \Comment{edge distribution policy} \label{line:distribution}\\

  \ForEach{edge in edgelist}{
    computeEndPoints(edge, src, dst, prefixWork) \Comment{use binary search} \label{line:binarysearch}\\
    applyOperator(g, src, dst, wl) \label{line:operator} \Comment{apply operator}
  }
}
\end{algorithm}

\subsection{Design}

ALB detects thread-block-level (CTA) load imbalance with minimal
overhead and efficiently balances the load among all threads of all thread
blocks. In graph analytical applications, CTA load imbalance occurs when some
CTAs process active vertices with \emph{huge} degrees while others process
active vertices with low degrees (Section~\ref{sec:twc}).  To detect imbalance,
ALB uses an inspection phase to separate high-degree active vertices from lower-degree
vertices that will not cause load imbalance with the TWC scheme (Algorithm~\ref{algo:inspect}).  
In the execution phase, ALB distributes the processing
of the high-degree active vertices across CTAs in its execution phase
(Algorithm~\ref{algo:execute}), eliminating load imbalance for such vertices,
and it uses the TWC scheme for the other vertices.

\noindent{\bf Inspection (Algorithm~\ref{algo:inspect}):}
Inspection iterates over active vertices
(Line~\ref{line:traverse}) and uses a lightweight check
(Line~\ref{line:degreeCheck}) to determine if a vertex is high degree based on
a threshold value. If so, it is pushed onto a separate worklist
(Line~\ref{line:push}).  If not, it is assigned to a TWC bin and
processed with TWC (Line~\ref{line:twc}).

\noindent{\bf Execution (Algorithm~\ref{algo:execute}):}
Each thread in each thread block (1) identifies edges that it processes
(Line~\ref{line:distribution}), (2) computes the endpoints of those edges
(Line~\ref{line:binarysearch}), and (3) applies the operator~\cite{pingali11}
along the edge (Line~\ref{line:operator}).  There are many distribution
policies for assigning edges to threads for Step (1). Step (2)
searches for the source (or destination) of the edge in the graph.

\begin{figure*}[t]
	\includegraphics[scale=0.31]{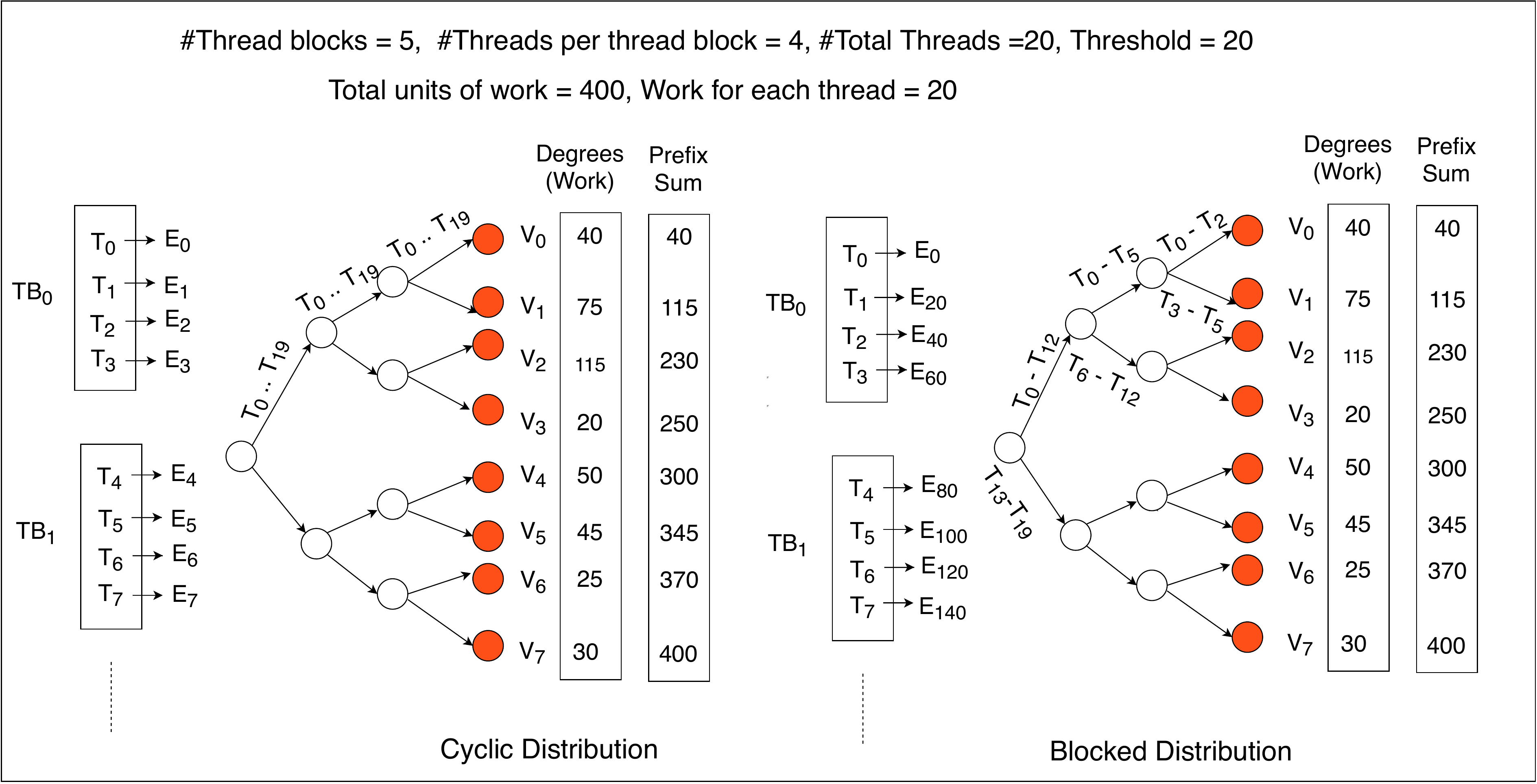}
  \caption{Cyclic and blocked edge distribution and binary search for
  vertices.  Distribution of edges shown in thread blocks while binary search
  is shown by trees.}
	\label{fig:distribution}
	\vskip -3mm
\end{figure*}

\emph{Distribution}:
We used two policies illustrated in Figure~\ref{fig:distribution}:
\emph{cyclic} and \emph{blocked}.  In cyclic distribution, edges are assigned
to threads in a round-robin manner, while in blocked distribution, each thread
is assigned a contiguous set of edges. More precisely, if the number of threads
is $p$ and the total number of edges of all huge vertices is $w$, thread $T_i$
is assigned edges $e_{i}, e_{p+i} \ldots e_{(w-1)*p+i}$ in cyclic and
$e_{(i*w)}, e_{(i*(w+1))} \ldots e_{(i+1)*w - 1}$ in blocked.

\emph{Search}:
If the graph is stored in COO format, edge endpoints are readily available, but
like most graph analytical systems, our system stores the graph in CSR (or CSC)
format to save space.  Therefore, to find the source (or destination) of an
edge, we perform a binary search (Line~\ref{line:binarysearch}) on
\emph{prefixWork}, which is the prefix sum of out-(or in-)degrees of the huge
vertices in \emph{work}.  For example, in Figure~\ref{fig:distribution}, thread
$T_4$ needs to process edge $e_4$ in the cyclic distribution, so it performs a
binary search in \emph{prefixWork} to find the source vertex $v_{0}$ (which has
the first 40 edges). The performance of binary search is sensitive to the
distribution. To illustrate in Figure~\ref{fig:distribution}, in cyclic
distribution, consecutive threads ($T_0 .. T_{19}$) process consecutive edges
($E_0 .. E_{19}$) whose binary search computations follow the same trajectory
in the binary search tree, leading to low thread divergence and good locality.
In blocked distribution, threads $T_0 ..  T_{19}$ compute source vertices by
following different paths in the binary search tree, leading to thread
divergence and poor locality.
Hence, we use cyclic distribution by default.

\noindent{\textbf{Threshold:}}
The threshold for classifying huge vertices can affect the performance of
applications. Setting it to 0 places all vertices in the huge bin;
this may be good for load balancing, but there will be overhead 
from the binary searches. Conversely, setting it to a value larger
than the maximum degree of vertices in the graph ensures that no vertex will be
placed in the huge bin; this eliminates binary search but hurts
load balance.
%
Determining the optimal threshold is difficult due to several unknown hardware
parameters, such as hardware thread block scheduling, warp scheduling, and
cache replacement policies.
However, setting the threshold to be equal to the number of threads launched in
a kernel ensures that the threads in each warp follow \emph{at most two
divergent branches} during the binary search with the cyclic distribution since
the threads will be assigned edges of \emph{at most two different vertices per
cyclic distribution of edges}.
Setting the threshold to less than the number of threads launched leads to more
divergent branches, resulting in thread divergence and poor locality.  Thus, we
set the threshold as the number of the threads  by default.

\subsection{Analysis} \label{sec:analysis}


\noindent{\textbf{Memory Accesses:}}
Table~\ref{tbl:notations}  shows notation used in analyzing the complexity of
the number of memory accesses in a computation round for cyclic and blocked
distribution policies.
We set the threshold $t$ to the number of threads $T$.

Each instance of a warp processes $W$ edges,  so each warp needs $e/(W*N_{w})$
instances to process all it edges.  Each divergent branch in a warp during
binary search requires $O(log(v))$ memory accesses.  We assume there is no
locality among the accesses generated by different warps (worst case).

\begin{wraptable}{r}{6.0cm}
\vspace{-15pt}
\footnotesize
\centering
\caption{Notations}
\label{tbl:notations}

\begin{tabular}{c|l}
\toprule
\multicolumn{1}{c|}{\textbf{Parameter}} & \multicolumn{1}{l}{\textbf{Description}} \\
\midrule
\textbf{$t$}                     & Threshold                    \\
\textbf{$T$}                      & Number of threads                      \\
\textbf{$W$}                     & Warp size                    \\
\textbf{$N_{w}$}                     & Number of Warps                    \\
\textbf{$e$}                      & Sum of degree of huge vertices                    \\
\textbf{$v$}                      & Number of huge vertices                    \\
\bottomrule
\end{tabular}
\vspace{-12pt}
\end{wraptable}

In cyclic distribution, threads in a warp processes successive edges.  As
the threads within a warp follow at most two divergent paths during binary
search,
the number of memory accesses in each warp
is $O(log(v))$.  Thus, the total number of memory accesses in all $N_w$ warps
is $O(e*log(v)/W)$.

In blocked distribution, each thread in a warp processes non-contiguous edges.
In the worst case, each thread can follow a different divergent branch.
Hence, the number of memory accesses in each warp
is $O(W * log(v))$.  Consequently,  the number of memory accesses in all $N_w$
warps is $O(e * log(v))$.

\noindent{\textbf{Memory Usage:}}
In each round, ALB uses two worklists to keep track of huge vertices and their
prefix sums, the latter of which reduces the overhead of binary search.
Although they incur $O(V)$ space overhead, in practice, $v \ll V$, where $v$
is the number of huge vertices.


\begin{figure}[t]
	\input{figs/sssp}
	\caption{A snippet of single-source shortest path (\sssp) program written in IrGL.}
	\label{fig:sssp}
\end{figure}

\subsection{Code Generation}

\noindent{\bf Single GPU:}
We implemented ALB in the IrGL compiler~\cite{irgl} to generate CUDA code
for a single GPU.  The compiler supports programming constructs to traverse
vertices and edges in parallel. Figure~\ref{fig:sssp} shows a snippet of the
\sssp{} program written in IrGL.
The outer loop
(Line~\ref{line:active_nodes}) iterates over the active vertices in each round,
and the inner loop (Line~\ref{line:neighbor}) processes the neighbors of the
active vertex. The \sssp{} operator (known as the relaxation operator)
is applied in each iteration of the inner loop, and it corresponds to a unit of
work (load) for this program.

For the \sssp{} program in Figure~\ref{fig:sssp}, the ALB code generated by our
modified IrGL compiler is shown in Figure~\ref{fig:sssp_lb}.
Lines~\ref{line:inspectBegin}-\ref{line:inspectEnd} show the generated code for
the inspection phase.  The function \emph{SSSP\_LB} shows the generated code
for the execution phase.  The main function
(Lines~\ref{line:highDegreeNodes}-\ref{line:lbKernel}) invokes the inspection
phase to identify the vertices in a huge bin (a worklist denoted as
\emph{work}).  In the presence of load imbalance, it invokes the execution
phase (Line~\ref{line:lbKernel}) after computing the prefix sum
(Line~\ref{line:prefixSum}) of all the vertices in the huge bin.

\begin{figure}[t]
	\input{figs/sssp_lb}
	\caption{A snippet of our compiler generated CUDA code for the
		\sssp{} program.}
	\label{fig:sssp_lb}
\end{figure}

\noindent{\bf Distributed GPUs:}
Most multi-GPU graph analytical systems use the
BSP execution model.  Hence, thread block imbalance within a GPU can lead
to stragglers and exacerbate load imbalance among GPUs.
To show the strengths of ALB, we
use ALB generated code with the CuSP~\cite{cusp} graph partitioner
and the Gluon~\cite{gluon} communication substrate (the Abelian~\cite{abelian}
compiler can generate Gluon communication code automatically).  CuSP partitions
the input graph using various partitioning policies selectable at runtime,
and Gluon manages synchronization of vertex labels. We denote the
resulting system D-IrGL (ALB).


\section{Evaluation Methodology}
\label{sec:setup}

\begin{table}[t]
\vspace{-15pt}
\footnotesize
\centering
\caption{Inputs and their key properties.}
\label{tbl:inputs}
\scalebox{0.8}{
\begin{tabular}{@{ }l@{\hskip1.0pt}||r@{\hskip4pt}r@{\hskip4pt}r@{\hskip4pt}r@{\hskip2pt}|r@{\hskip4pt}r@{\hskip4pt}r@{\hskip4pt}r@{ }}
\toprule
                                          &        &  \textbf{rmat25} & \textbf{orkut}      & \textbf{road-USA}                  & \textbf{rmat26}   & \textbf{rmat27}  & \textbf{twitter40}   & \textbf{uk2007}  \\
\midrule                                                                                                                                                                                                                              
\textsf{$|V|$}                            &                   &  33.5M           & 3.1M                & 23.9M                        &    67.1M               & 134M             &   41.6M                & 106M              \\
\textsf{$|E|$}                            &                 &  536.8M          & 234M                & 57.7M                        &    1,074M              & 2,147M           &   1,468M               & 3,739M             \\
\textsf{$|E|/|V|$}                        &                   &  16              & 76                  & 2                            &                   & 16               &   35                   & 35                 \\
\textsf{max $D_{out}$}                    &                  &  125.7M           & 33,313              & 9                            &    239M                & 453M             &   2.99M                & 15,402              \\
\textsf{max $D_{in}$}                     &                &  14733           & 33,313              & 9                            &    18211               & 21806            &   0.77M                & 975,418             \\
\textsf{Approx.}                          &                    & 3                & 6                   & 6261                         &      3                 & 3                &    12                  & 115                 \\
\textsf{Diameter}                         &                       &                   &                      &                              &                        &                  &                      &                  \\
\textsf{Size (GB)}                        &                    & 4.3              & 1.8                 & 0.6                          &     8.6                & 18               &    12                  & 29                 \\
\bottomrule

\end{tabular}}
\vspace{-10pt}
\end{table}

For our evaluation, we used 2 different setups on the Bridges supercomputer.
For single-GPU and single-host multi-GPU experiments, we used a single machine
(this setup is referred to as \emph{Bridges-Volta}) on the Bridges
supercomputer with 2 Intel Xeon Gold 6148 CPUs (with 128GB RAM) and 8 NVIDIA
Volta V100 GPUs each with 16GB of memory.  For multi-host multi-GPU
experiments, we used multiple machines (this setup is referred to as
\emph{Bridges-Pascal}) on the Bridges supercomputer each with 2 Intel Broadwell
E5-2683 v4 CPUs (with 128GB RAM) and 2 NVIDIA Tesla P100 GPUs each with 16GB of
memory.  The network interconnect is Intel Omni-Path. All applications were
compiled using CUDA 9.0, gcc 6.3.0, and MVAPICH2-2.3. Threshold for {\it huge}
vertices in ALB is the number of threads launched for the application kernel
(based on GPU architecture): $163840$ and $114688$ on Bridges-Volta and
Bridges-Pascal, respectively.

Table~\ref{tbl:inputs} lists the input graphs used; 
the table splits the graphs
into {\it small} graphs that are evaluated on Bridges-Volta (single machine) 
and {\it large} graphs that are evaluated on Bridges-Pascal (multiple machines). 
We run on up to 16
GPUs (8 machines). 

We evaluated five applications: breadth-first search (bfs), connected components
(cc), k-core decomposition (kcore), pagerank (pr), and single-source shortest
path (sssp). 
All applications are run until convergence.  The reported execution time is an 
average of three runs excluding graph 
construction time.


For single-host multi-GPU experiments, we compare with Gunrock~\cite{gunrock},
and for multi-host multi-GPU experiments, we compare with D-IrGL~\cite{gluon}
and Lux~\cite{lux}.  Gunrock is single host multi-GPU graph analytical
framework and supports TWC and LB load balancing.
For Gunrock, we used bfs, sssp, and cc; it does not have kcore, and we omit pr
as it did not produce correct results.  Gunrock provides bfs with and without
direction optimization. The other systems do not have direction optimization;
therefore, we evaluated bfs in Gunrock without direction optimization.  D-IrGL
and Lux are a distributed multi-GPU graph analytical frameworks.  D-IrGL uses
TWC load balancing. Applications in D-IrGL use push operators (traverse
outgoing edges) except for pr and kcore which use pull operators (traverse
incoming edges).
Lux uses a variant of TWC 
load balancing.  For
Lux, we use only cc and pr: the other applications are either not available or
not correct.

\section{Experimental Results}
\label{sec:results}


\subsection{Performance Analysis on a Single GPU Platform} \label{sec:singlegpu}

Table~\ref{tbl:singleGPU} compares the performance of our adaptive load
balancing approach, D-IrGL (ALB), 
with D-IrGL (TWC), D-IrGL (LB), Gunrock
(TWC), and Gunrock (LB).  
D-IrGL (TWC) is default D-IrGL. 
D-IrGL (LB) is D-IrGL (ALB) without adaptive balancing: it balances all active
vertices regardless of their degrees. Gunrock (TWC) uses TWC, and Gunrock (LB)
uses its LB. 



\begin{wraptable}{R}{8.4cm}
\vspace{-15pt}
\footnotesize
\centering
\caption{Execution time (ms) on a single V100 GPU. }
\label{tbl:singleGPU}
\scalebox{0.9}{\begin{tabular}{l|l|rrrrr}
\toprule
\multicolumn{1}{c|}{\textbf{Input}} & \multicolumn{1}{c|}{\textbf{App}} & \multicolumn{1}{c}{\textbf{Gunrock}} & \multicolumn{1}{c}{\textbf{Gunrock}} & \multicolumn{1}{c}{\textbf{D-IrGL}} & \multicolumn{1}{c}{\textbf{D-IrGL}} & \multicolumn{1}{c}{\textbf{D-IrGL}} \\
 & & \multicolumn{1}{c}{\textbf{(TWC)}} & \multicolumn{1}{c}{\textbf{(LB)}} & \multicolumn{1}{c}{\textbf{(TWC)}} & \multicolumn{1}{c}{\textbf{(LB)}} &  \multicolumn{1}{c}{\textbf{(ALB)}}\\
\midrule
\multirow{5}{*}{rmat25}            & bfs                              & 1419.6                                    & 299.9                                   & 1015.7                                     & 120.8                                     & \colorbox{blue!20}{\textbf{113.7}}                                    \\
                                    & sssp                             & 1321.2                                    & 346.3                                    & 1418.0                                    & 150.2                                      & \colorbox{blue!20}{\textbf{142.4}}                                    \\
                                    & cc                               & 347.7                                    & 382.3                                   & 648.3                                    & 265.3                                      & \colorbox{blue!20}{\textbf{142.4}}                                      \\
                                    & pr                               & -                                         & -                                        & \colorbox{blue!20}{\textbf{1418.0}}                                  & 3948.4                                     & 1423.0                                  \\
                                    & kcore                            & -                                         & -                                        & 1561.3                                   & 317.5                                     & \colorbox{blue!20}{\textbf{247.6}}                                     \\
\midrule
\multirow{5}{*}{orkut}             & bfs                              & 62.2                                     & 36.5                                   & 17.7                                    & \colorbox{blue!20}{\textbf{16.6}}                                     & 18.4                                      \\
                                    & sssp                             & 166.4                                     & 137.1                                    & \colorbox{blue!20}{\textbf{55.3}}                                    & 67.5                                     & 57.3                                    \\
                                    & cc                               & \colorbox{blue!20}{\textbf{22.9}}                                     & 23.5                                    & 28.0                                    & 35.8                                     & 29                                    \\
                                    & pr                               & -                                         & -                                        & \colorbox{blue!20}{\textbf{1562.7}}                                    & 2889.3                                     & 1578.6                                  \\
                                    & kcore                            & -                                         & -                                        & \colorbox{blue!20}{\textbf{86.0}}                                      & 154.2                                     & 90.6                                    \\
\midrule
\multirow{5}{*}{road-USA}          & bfs                              & 355.7                                     & \colorbox{blue!20}{\textbf{335.7}}                            & 2287                                  & 4266.7                                      & 2531.0                                  \\
                                    & sssp                             & 20883.6                                   & 20497.2                                  & \colorbox{blue!20}{\textbf{6436.0}}                                    & 16465.3                                     & 8902.0                                    \\
                                    & cc                               & 31.3                                     & \colorbox{blue!20}{\textbf{33.7}}                                    & 3066.0                                  & - 	& 3300.7                                                                       \\
                                    & pr                               & -                                         & -                                        & \colorbox{blue!20}{\textbf{450.0}}                                   & 877.2                                     & 458.5                                   \\
                                    & kcore                            & -                                         & -                                        & \colorbox{blue!20}{\textbf{3.0}}                                       & 14.4                                     & 3.4                                      \\
\bottomrule
\end{tabular}}
\vskip -4mm 
\end{wraptable}

\noindent{\textbf{D-IrGL and ALB:}}
D-IrGL (ALB) is $4.6\times$ faster on average compared to D-IrGL (TWC) for bfs,
sssp, cc, and kcore on rmat25.  These application and input configurations
have heavy load imbalance across thread blocks in some iterations, and
ALB detects this and balances the workload. D-IrGL (ALB) does not
improve pr runtime on rmat25 because pr iterates over the incoming neighbors of a
vertex as opposed to the other applications which iterate over outgoing neighbors.  The
in-degree distribution ($D_{in}$, Table~\ref{tbl:inputs}) is not as skewed as the out-degree
distribution ($D_{out}$), so large imbalance does not exist. Similar reasoning
explains why D-IrGL (ALB) does not outperform D-IrGL (TWC) for applications on
the other graphs: both road-USA and orkut have low load imbalance (the skew of
in and out degrees is relatively small), so ALB does not detect huge
vertices and thread block load imbalance. 

D-IrGL (LB) illustrates the importance of selectively balancing only
high-degree (huge) vertices: D-IrGL (LB) performs worse than D-IrGL (ALB) for
most configurations. Even though D-IrGL (LB) addresses the load imbalance for
rmat25, it suffers from the overhead of balancing all vertices in every
iteration for other configurations. In particular, it significantly
underperforms on road-USA: since road-USA does not have imbalance,
binary search overhead from balancing leads to worse performance.





\begin{figure*}[t]
	\centering
	\subfloat[bfs and pr using TWC\label{fig:no_lb_pr_bfs_rmat25}]{%
		\includegraphics[scale=0.27]{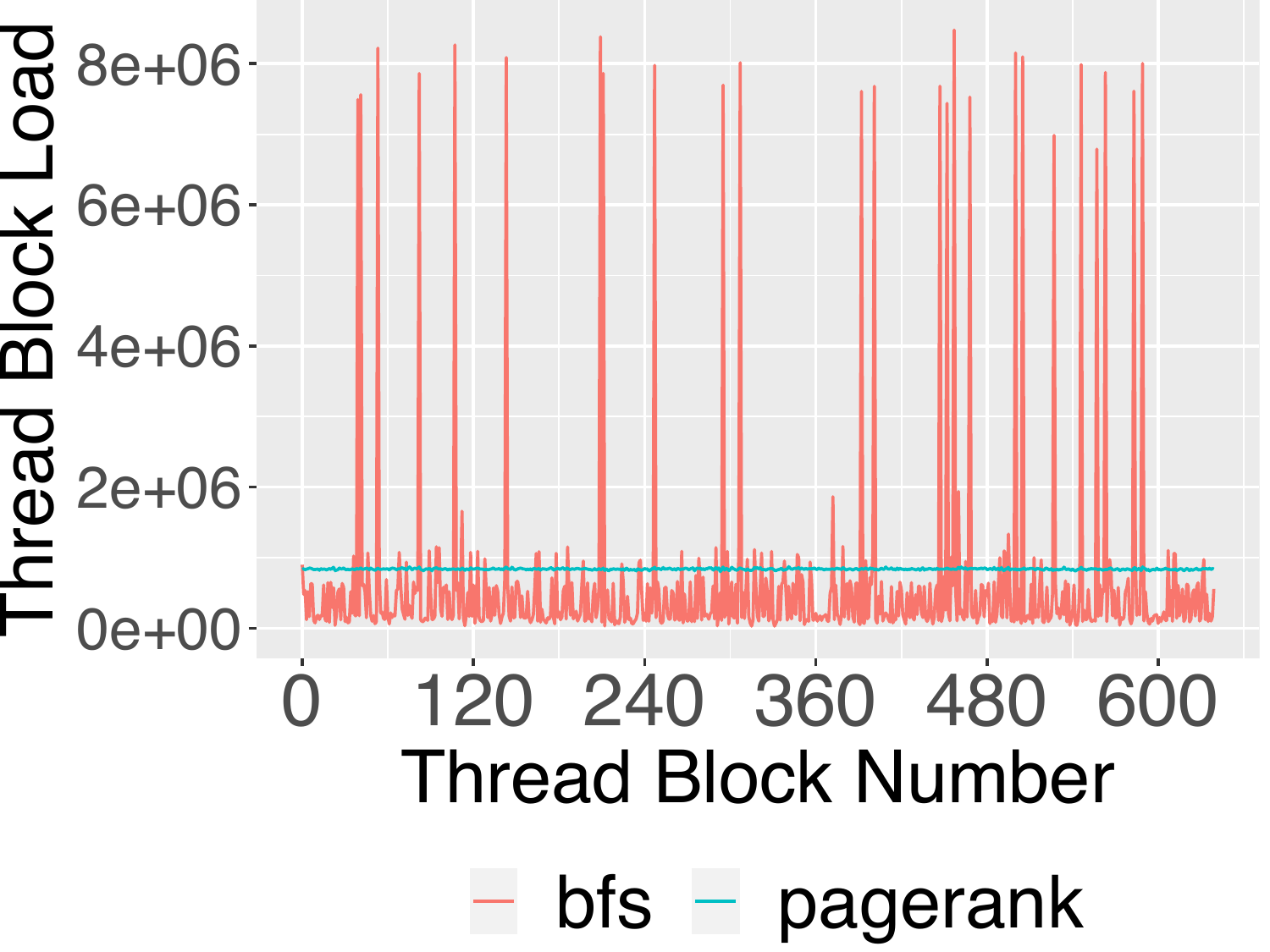}        
	}
	\subfloat[bfs using ALB\label{fig:lb_bfs_rmat25}]{%
		\includegraphics[scale=0.27]{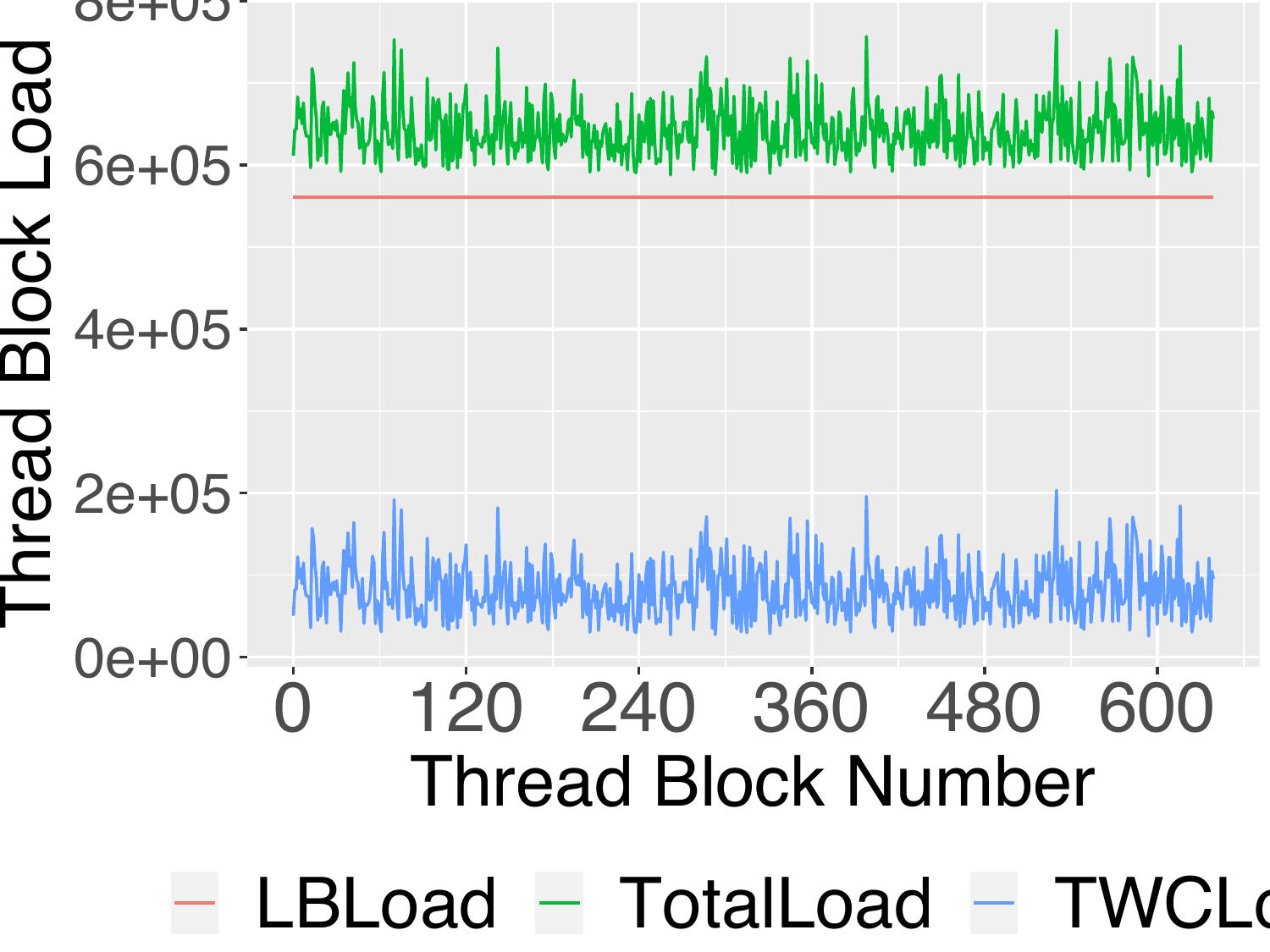}
	}
	\subfloat[pr using ALB\label{fig:lb_pr_rmat25}]{%
		\includegraphics[scale=0.27]{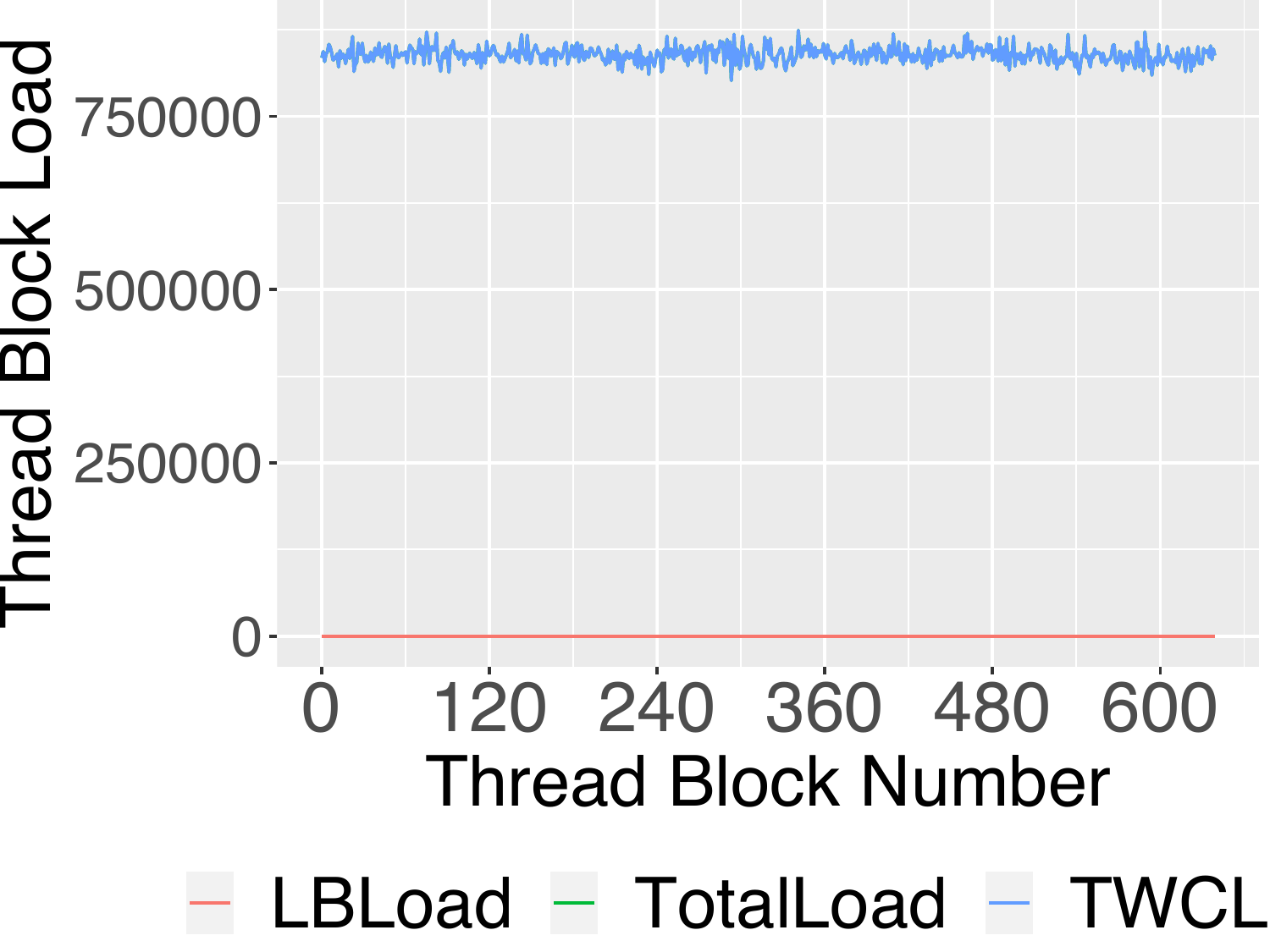}        
	} 
	\caption{Thread block load distribution of D-IrGL for rmat25 on 1 V100 GPU.}
	\vskip -4mm 
\end{figure*}


\noindent{\textbf{Gunrock and ALB:}}
Gunrock (LB) generally performs better than Gunrock (TWC) due to thread block
load imbalance in TWC. D-IrGL (ALB) outperforms both LB 
and TWC in Gunrock for 
most of the applications on rmat25 and orkut 
due to its adaptivity: it fixes
imbalance if it exists, and if a round does not have imbalance, ALB has minimal
load balancing overhead.


Gunrock (LB) performs better than D-IrGL (ALB) for bfs and cc on road-USA.
Gunrock uses an explicit sparse work-list to maintain the active vertices,
whereas D-IrGL use an implicit dense work-list 
(boolean vector). As bfs and cc have few active
vertices in a round of computation, implicit worklists do not perform well: 
finding active vertices requires iterating over the entire worklisti.


\noindent{\textbf{Thread Block Load Distribution:}}
We examine the thread block load distribution of work with 
TWC and with ALB to
better understand why ALB performs well.  Figure~\ref{fig:no_lb_pr_bfs_rmat25} shows
the thread block work distribution for D-IrGL (TWC) for bfs and pr on rmat25. 
Figure~\ref{fig:lb_bfs_rmat25} and Figure~\ref{fig:lb_pr_rmat25} show the
distribution with D-IrGL (ALB) for bfs and pr, respectively.  The figures show
the distributions for two kernels launched in our approach: (1) LB, which
distributes the work of the huge active 
vertices among all the thread blocks, and
(2) TWC, which distributes the load of all other active vertices to
threads/warp/CTA based on their degrees.  The figure also shows the total load
(the sum of TWC and LB).

D-IrGL (TWC) has heavy load imbalance for bfs.  D-IrGL (ALB), however,
has a more balanced load distribution than D-IrGL (TWC) because ALB 
distributes the load of high-degree active 
vertices equally among all the thread
blocks.  As pr does not have from thread block load imbalance in D-IrGL, it
performs well for both D-IrGL (TLB) and D-IrGL (ALB). Notably, ALB does not
instantiate the LB kernel as it detects that no imbalance exists.




\begin{wrapfigure}{r}{4.7cm}
\vspace{-10pt}
\centering
\includegraphics[page=1,width=0.4\textwidth]{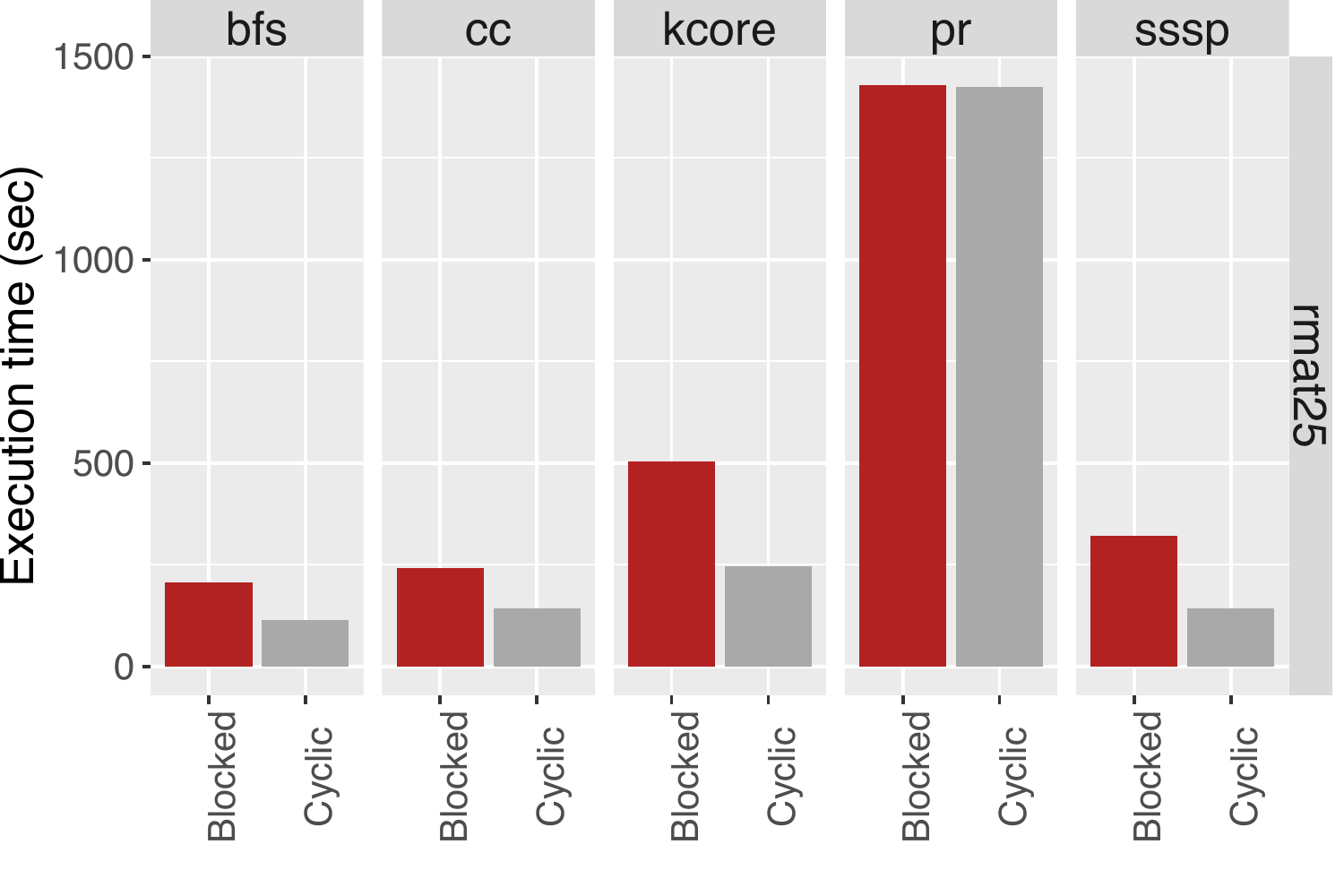}
\caption{D-IrGL (ALB) with cyclic and blocked distribution for rmat25 on 1 V100 GPU.}
\label{fig:cyclic}
\vspace{-10pt}
\end{wrapfigure}

  


\noindent{\textbf{Analyzing Threshold:}}
The threshold for a high-degree (huge) vertex in our experiments was the thread count,
but to explore the effect of the threshold on performance,
we varied the
threshold value by powers of 2 
starting from 1 (including 0 as well) for pr and sssp on road-USA and rmat25.
We found that for configurations with low imbalance (pr on both
graphs and both applications on road-USA), any threshold value starting from $2^8$ (the
thread block size) performed similarly, 
while for the configuration with
imbalance (sssp, rmat25), 
execution time suffered when the threshold was made
larger than the max degree 
(making it equivalent to TWC as no vertex will be 
considered huge). 
For both inputs, which have widely different 
characteristics, 
ALB performed similarly for 
any threshold in the range of $2^{8}$ to $2^{18}$. 
In other words, there are many good thresholds, 
and whether a threshold value is good or not 
depends more on the architecture than on the 
input or application.

\noindent{\textbf{Cyclic vs. Blocked Distribution:}}
Figure~\ref{fig:cyclic} compares the performance of cyclic and blocked
distribution for D-IrGL (ALB) for rmat25.  Cyclic distribution outperforms
blocked distribution for all configurations (except pr), and it is on average 
$1.7\times$ faster.  As explained in Section~\ref{sec:lb}, cyclic
distribution has better locality and benefits from lower thread divergence
compared to blocked distribution.

\begin{figure*}[t]
	\subfloat[\scriptsize{Bridges-Volta (V100 GPUs)}\label{fig:singleHostMultiGPUs}]{%
		\centering
		\includegraphics[page=1,width=0.49\textwidth]{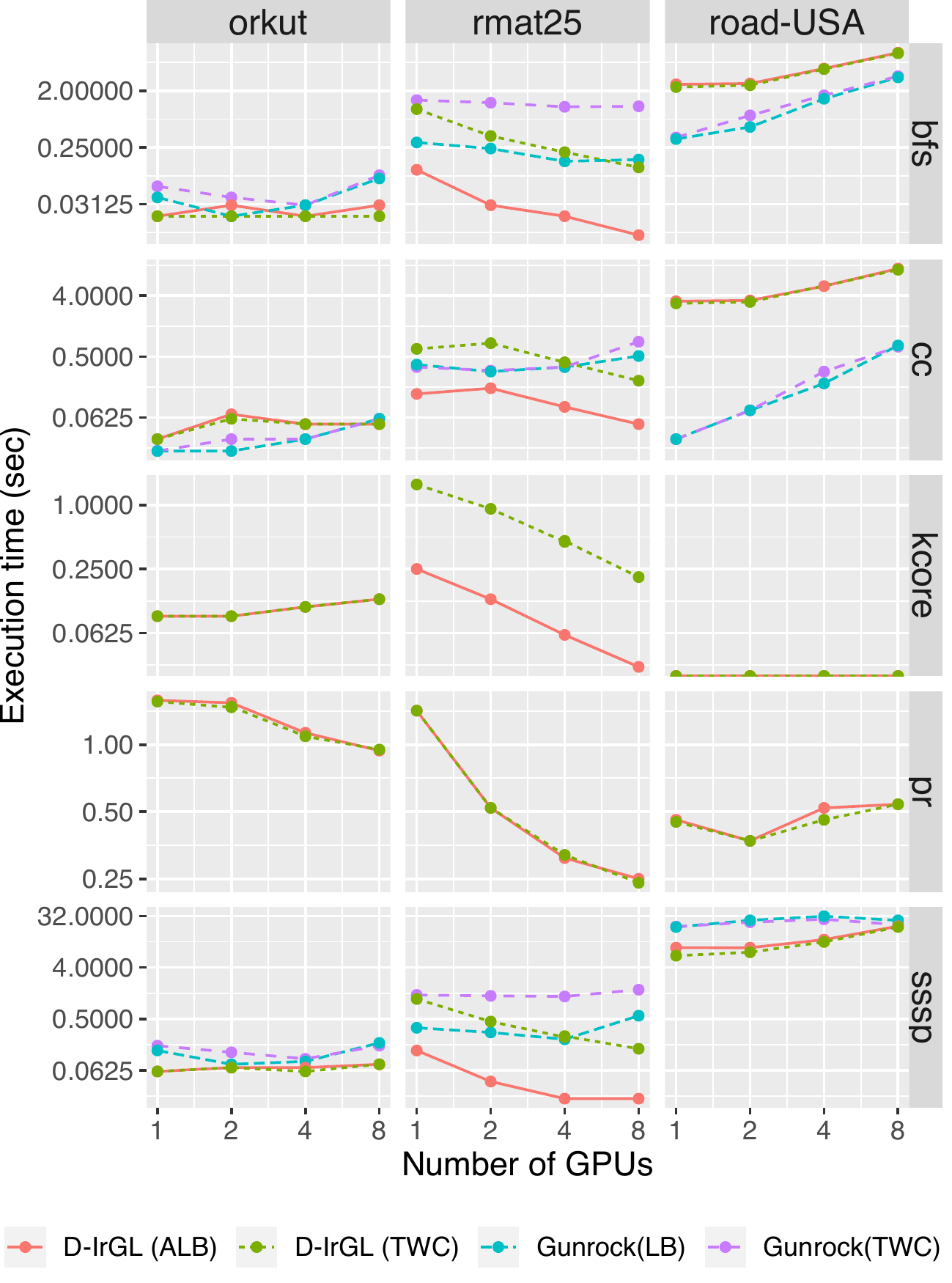}
	} 
	\subfloat[\scriptsize{Bridges-Pascal (P100 GPUs)\label{fig:bridges}}]{%
		\centering
		\includegraphics[page=1,width=0.49\textwidth]{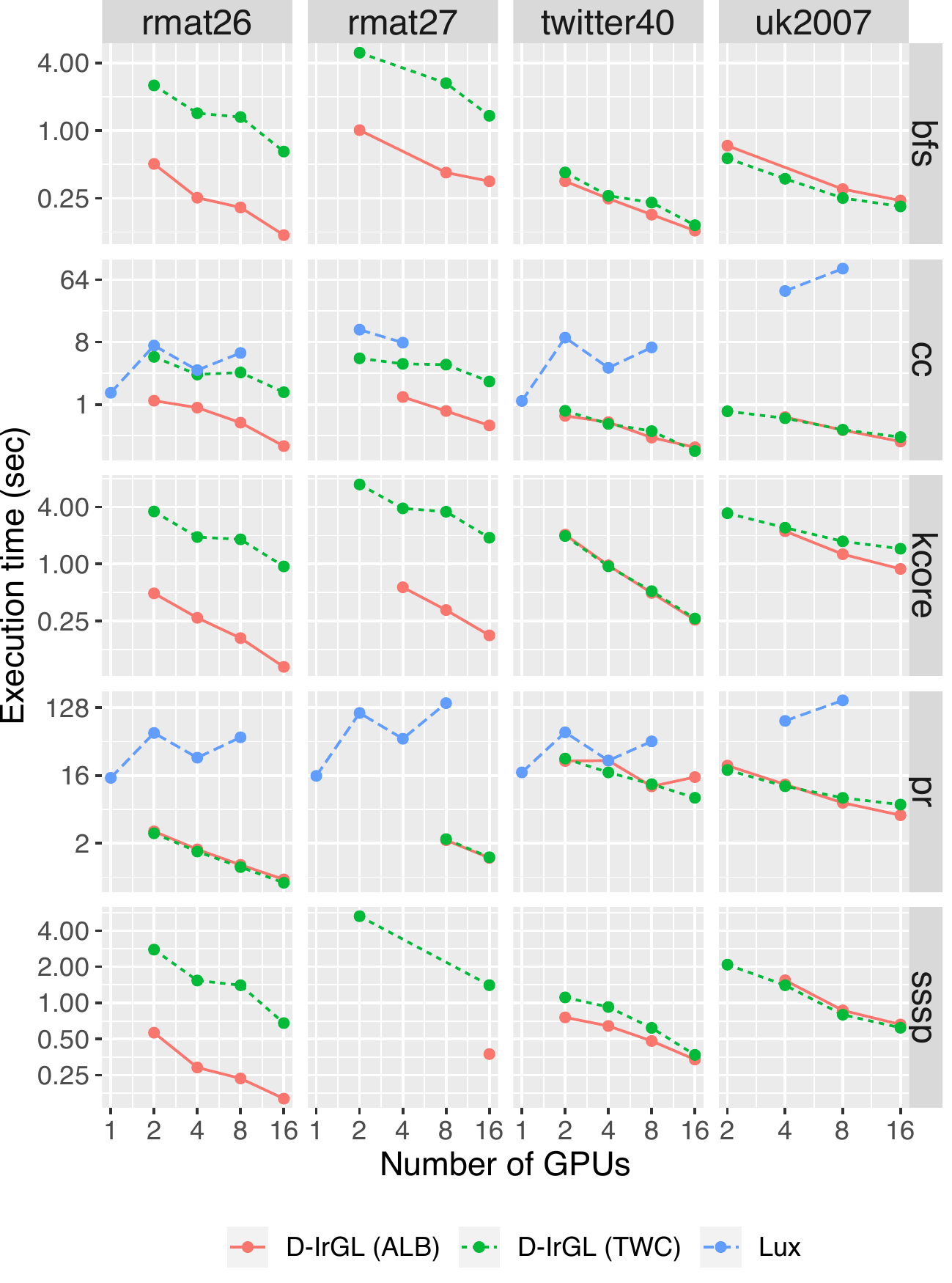}
	}
	
	\caption{Strong scaling of different systems on multi-GPU platforms.}
\end{figure*}

\subsection{Performance Analysis on Multi-GPU Platforms} \label{sec:singleHostMultiGPUs}

\noindent{\bf Single-Host Multi-GPU:}
Figure~\ref{fig:singleHostMultiGPUs} shows strong scaling of D-IrGL (TWC),
D-IrGL (ALB), Gunrock (TWC), and Gunrock (LB) for small graphs on
Bridges-Volta.  The trends are similar to those on 1 Volta GPU.  For all
applications on orkut and road-USA and for pr on rmat25, D-IrGL (ALB) and
D-IrGL (TWC) perform similarly because ALB's inspection phase does not detect
thread block load imbalance (i.e., huge vertices).  Both are also better than
Gunrock in most cases.  For bfs, cc, kcore, and sssp on rmat25, D-IrGL (ALB)
outperforms D-IrGL (TWC), Gunrock (TWC), and Gunrock (LB).  ALB
improves computation on each GPU, thereby reducing the total execution time. 


\begin{wrapfigure}{R}{4.4cm}
\vspace{-10pt}
		\centering
		\includegraphics[page=1,width=0.38\textwidth]{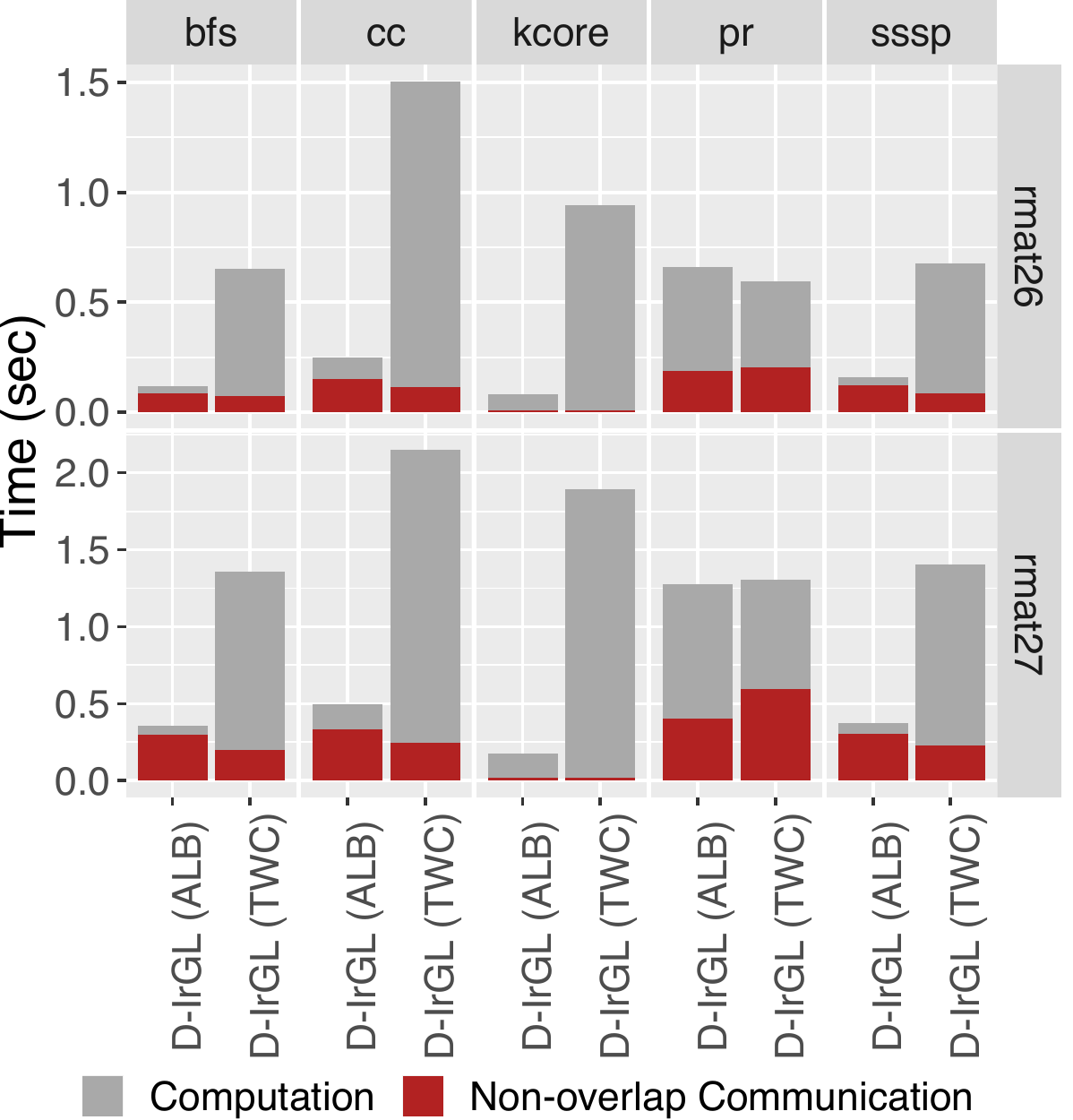}
		\caption{Execution time breakdown of D-IrGL on 16 P100 GPUs.\label{fig:bridgesBreakdown}}
		\label{fig:part}
\vspace{-10pt}
\end{wrapfigure}

\noindent{\bf Multi-Host Multi-GPU:}
Figure~\ref{fig:bridges} shows the strong scaling of D-IrGL (ALB), D-IrGL
(TWC), and Lux for large graphs on Bridges-Pascal (we do not evaluate Gunrock
because it is restricted to a single-host).  Both D-IrGL (ALB) and D-IrGL (TWC)
perform better than Lux for all configurations.  The only difference between
ALB and TWC is in computation on each GPU, and D-IrGL (ALB) performs similar to
or better than D-IrGL (TWC).

D-IrGL (ALB) performs slightly better than D-IrGL (TWC) for pr on uk2007, but
for the other applications on uk2007, they perform similarly.  uk2007's max
out-degree $15402$ is smaller than the threshold $114688$, so for push-style
algorithms like bfs, cc, kcore, and sssp, ALB does not detect any huge
vertices.  uk2007's max in-degree $975418$ is much higher, so ALB detects
the imbalance and corrects it.  For twitter40, D-IrGL (ALB) is
slightly faster than D-IrGL (TWC) on twitter40 in most cases because
twitter40's max out-degree and max in-degree are much higher than the
threshold. The max in-degree of rmat26 and rmat27 is smaller than the
threshold, so D-IrGL (ALB) and D-IrGL (TWC) perform similarly for pr.
For the other applications on rmat26 and rmat27, ALB detects and addresses
thread block load imbalance because their max out-degree is much higher than
the threshold.  In these cases, D-IrGL (ALB) is faster than D-IrGL (TWC) by
$4.3\times$ on average.


\noindent{\bf Analysis:}
To analyze the performance improvement for rmat26 and rmat27, we show the
breakdown of the total execution time into computation and 
non-overlapping communication time on 16 GPUs in
Figure~\ref{fig:bridgesBreakdown}.  The computation time of an application is
the time spent in executing the kernels of the application in the GPU.  We
report computation time as the maximum time among all GPUs.  Thus, computation
time acccounts for load imbalance.  The rest of the execution time is the
non-overlapping communication time, including the time to synchronize
vertex labels among the GPUs. 

The results show that most applications in D-IrGL (TWC) spend most of
the execution time in computation.  rmat26 and rmat27 have a very large max
out-degree ($D_{out}$ shown in Table~\ref{tbl:inputs}), so push-style
applications in D-IrGL (TWC) suffer from thread block load imbalance on one of
the GPUs. D-IrGL (ALB) reduces the computation time on the
(straggler) GPUs by balancing load across the thread blocks on each GPU. This
in turn balances the load among the GPUs, thereby reducing the total execution
time.

\noindent{\bf Summary:} Our adaptive load balancer (ALB) improves application
performance significantly on configurations with thread block load
imbalance on both single GPU and multi-GPU platforms and incurs minimal
overhead on configurations that have a balanced thread block load.



\section{Related Work}
\label{sec:related}

\noindent{\textbf{GPU Graph Analytics:}}
GPUs have been widely adopted for graph analytics, spanning single-host
single-GPU systems~\cite{irgl,sepgraph,autotuner}, CPU-GPU heterogeneous
systems~\cite{abelian}, single-host multi-GPU systems~\cite{groute,gunrock},
and multi-host multi-GPU systems~\cite{gluon,lux}.  This paper presents an
adaptive load balancer that can be incorporated in all these systems 
to improve the performance on each GPU. 


\noindent{\textbf{Load Balancing for GPU Graph Analytics:}}
%
Prior load-balancing schemes~\cite{nasre13datatopology,merrill12,enterprisegpu,gunrock} 
were described in Section~\ref{sec:challenges}. 
Unlike the vertex-based load-balancing scheme~\cite{nasre13datatopology} and TWC~\cite{merrill12}, 
ALB balances load across thread blocks. 
In contrast to edge-based load-balancing schemes~\cite{bcgpus}, 
ALB has lower memory overhead as it uses CSR format instead of COO format. 
Enterprise~\cite{enterprisegpu} is restricted to bfs and has 
overhead from barriers;
ALB uses only a constant number of barriers 
(one per bin)
to balance the load any graph application.
Gunrock's LB~\cite{gunrock} has overhead from many
binary searches because 
it distributes edges of all vertices among threads and 
uses a block distribution. 
ALB dynamically chooses vertices that 
would benefit from edge distribution among threads and 
uses a cyclic distribution to minimize binary search overhead.



\section{Conclusion}
\label{sec:conclusion}

We presented an adaptive load balancing mechanism for GPUs that detects the
presence of thread block load imbalance and distributes load equally among
thread blocks at runtime.  We implemented our strategy in the IrGL compiler
and evaluated its effectiveness using up to 16 GPUs.  Our approach
improves performance of applications by $2.2\times$ on average for
inputs with load imbalance compared to the previous load-balancing
schemes.

%
%
%
\bibliographystyle{splncs04}
\bibliography{References/related,References/iss,References/numa,References/graphs,References/gpugraphs,References/outofcore,References/partitioning,References/resilience,References/others}

\end{document}